# Zitterbewegung in the vicinity of the Chern-Simons black hole

NATALIA N. KONOBEEVA

*Volgograd State University, 400062, Volgograd, Russia*
yana_nn@inbox.ru

MIKHAIL B. BELONENKO

*Volgograd State University, 400062, Volgograd, Russia*
mbelonenko@yandex.ru

We consider zitterbewegung (ZB) effect in four and five-dimensional space in the vicinity of the Chern-Simons black hole with torsion. The metric is taken in the traditional spherically symmetric form. We consider the equation for the current in the framework of the Schrödinger representation. Dependences of the electric current density on time is calculated for different dimension of space. In particular, it is shown that, with an increase in the space dimension, the effect of trembling motion qualitatively coincides. As the dimension increases, the amplitude of this effect decreases. Also, we investigate the dependence of the current on the radial coordinate for different dimensions of space. We obtain that when away from the event horizon the current ZB increases.

**Acknowledgments**

This work was supported by the Ministry for Science and Higher Education of the Russian Federation under the government task No. 2.852.2017/4.6.

**1. Introduction**

Together with the great success of the general theory of relativity [1] from the beginning of the emergence of this theory, its various generalizations arise. In particular, the great interest of researchers is the consideration of torsion in the generalized theory. There are many theories that take into account torsion, for example, Poincaré's gauge gravity theory, tele-parallel gravity and so on. The simplest generalization, apparently, is Einstein-Cartan's theory of gravity [2, 3]. In it, the curvature and torsion are independent variables, and the source for torsion is the spin degrees of freedom of material fields.

It is well-known that the Einstein equations in the absence of matter have a non-trivial solution (the so-called Einstein space). At the same time, it is shown that in four dimensions the torsion is zero in the absence of sources. It should be noted that with an increase in the dimension of the space [4], the torsion equations admit nontrivial solutions in the absence of sources. These solutions become new nontrivial degrees of freedom [5].

Some space-time solutions for black holes with torsion in these theories are studied in [6-8], including exact ones [9]. Three-dimensional torsional gravity was also considered in [10, 11], where supersymmetric expansion in Chern-Simons formulation was investigated. This model allows as a solution a generalization of the BTZ black hole (Bañados, Teitelboim, Zanelli) with torsion [12]. There are a number of papers related to the inclusion of torsion in the geometry in the context of AdS/CFT compliance [7, 13, 14].

Note that the inclusion of torsion causes new physical effects and in 2+1 dimensions is associated with the continual theory of lattice defects in solid state physics [15].

On the other hand, interest in the ZB effect (trembling motion), including in the vicinity of the black hole, has recently increased [16, 17]. It is believed that ZB can be observed in any system when the velocity operator does not commute with the Hamiltonian [18].

Considering the above-mentioned about the connection of torsion with the theory of defects, there is a problem about the ZB effect consideration, which has already been discovered experimentally [19], in torsion-based theories.

## 2. Basic equations

The metric for the Chern- Simons black hole is [20]:

$$ds^2 = -f(r)^2 dt^2 + \frac{1}{f(r)^2} dr^2 + r^2 d\sigma_\gamma^2,$$

$$f(r)^2 = \frac{r^2}{l^2} - \mu, \mu = \alpha(2\sigma G_k)^{1/k} - \gamma \tag{1}$$

with the horizon located near: $r_+ = l\sqrt{\mu}$.

The minimally coupled fermion field with curvature against a Chern-Simons black hole in d-dimensions is given by the Dirac equation:

$$(\gamma^\mu \nabla_\mu + m)\Psi = 0,$$

$$\nabla_\mu = \partial_\mu + \Gamma_\mu, \tag{2}$$

$$\Gamma_\mu = \frac{1}{4}\gamma^a \gamma^b e_a^\nu \left[\partial_\mu e_{b\nu} - \Gamma^\sigma_{\mu\nu} e_{b\sigma}\right]$$

Gamma matrices in curved space are defined as: $\gamma^\mu = e_a^\mu \gamma^a$, where $\gamma^a$ are the gamma matrices in flat space, $e_a^\mu$ are the fielbein fields, $\Gamma^\sigma_{\mu\nu}$ are second order Christoffel symbols.

Using the substitution:

$$\Psi_1 \pm \Psi_2 = \sqrt{\frac{\cosh\rho \pm \sinh\rho}{\sinh\rho \cdot \cosh^{d-2}\rho}} (\Psi'_1 + \Psi'_2) \qquad (3)$$

as well as changing variables $x = \tanh^2\rho$ for our metric, the Dirac equation takes the form:

$$2(1-x)x^{0.5}\partial_x\Psi'_1 + i\left(\frac{\omega l}{\sqrt{\mu}}x^{-0.5} - \frac{\xi}{\sqrt{\mu}}x^{0.5}\right)\Psi'_1 + \left(i\frac{\omega l - \xi}{\sqrt{\mu}} + 0.5 + ml\right)\Psi'_2 = 0,$$

$$2(1-x)x^{0.5}\partial_x\Psi'_2 - i\left(\frac{\omega l}{\sqrt{\mu}}x^{-0.5} - \frac{\xi}{\sqrt{\mu}}x^{0.5}\right)\Psi'_2 + \left(-i\frac{\omega l - \xi}{\sqrt{\mu}} + 0.5 + ml\right)\Psi'_1 = 0 \qquad (4)$$

here $i\xi$ are the eigenvalues of the Dirac operator.

Separating the system of equations (4) and using the notation:

$$\Psi'_1 = x^\alpha (1-x)^\beta F(x),$$

$$\alpha = -\frac{i\omega l}{2\sqrt{\mu}}, \quad \beta = -0.5(0.5 + ml)$$

we obtain the equation for $F(x)$:

$$x(1-x)F''(x) + (c - (1+a+b)x)F'(x) - abF(x) = 0 \qquad (5)$$

which has the following solution:

$$\Psi'_1 = C_1 x^\alpha (1-x)^\beta {}_2F_1(a,b,c,x) + C_2 x^{0.5-\alpha}(1-x)^\beta {}_2F_1(a-c, b-c+1, 2-c, x) \qquad (6)$$

where ${}_2F_1(a,b,c,y)$ is the hypergeometric function.

If $C_2=0$, functions $\Psi'_1, \Psi'_2$ have the following form:

$$\Psi'_1 = C_1 x^\alpha (1-x)^\beta {}_2F_1(a,b,c,x),$$

$$\Psi'_1 = \frac{a-c}{c} C_1 x^{\alpha+0.5}(1-x)^\beta {}_2F_1(a, b+1, c+1, x) \qquad (7)$$

The eigenfunctions can be easily found from (3).

The Hamiltonian can be found from the Dirac equation:

$$H = ip \begin{pmatrix} ml + 0.5 + 2(1-x)\partial_x & \frac{ml+0.5}{x^{0.5}} - i\frac{\varepsilon x^{0.5}}{p \cdot l} + 2(1-x)x^{0.5}\partial_x \\ -\frac{ml+0.5}{x^{0.5}} - i\frac{\varepsilon x^{0.5}}{p \cdot l} - 2(1-x)x^{0.5}\partial_x & -ml - 0.5 - 2(1-x)\partial_x \end{pmatrix} \qquad (8)$$

$$p = \frac{\sqrt{\mu}}{l(x^{-1}-1)}$$

Let's go to the Schrodinger picture [21]:

$$\hat{r} = \hat{r}\big|_0, \; |\Psi\rangle = |\Psi(t)\rangle \tag{9}$$

and calculate the current according to the formula:

$$j = \langle \Psi(t)|\dot{r}|\Psi(t)\rangle \tag{10}$$

Since $i\dot{\Psi} = H\Psi$ or in the eigenvalues representation: $|\Psi(t)\rangle = \sum_\alpha C_\alpha e^{i\varepsilon_\alpha t}|\Psi(t)\rangle_\alpha\big|_0$, it is easy to obtain:

$$j = -i\langle \Psi(t)|[r,H]|\Psi(t)\rangle$$

And

$$\Psi(t) = \tilde{C}_1 \begin{pmatrix} \Psi_1(E) \\ \Psi_2(E) \end{pmatrix} e^{iEt} + \tilde{C}_2 \begin{pmatrix} \Psi_1(-E) \\ \Psi_2(-E) \end{pmatrix} e^{-iEt},$$
$$\tilde{C}_1, \tilde{C}_2 = const \tag{11}$$

Finally, the expression for the current has the form:

$$j(r,t) = \int dE \, e^{-\frac{(E-E_0)^2}{\Delta^2}} 2p(1-x) \Big( |\tilde{C}_1|^2 \big( -|\Psi_1(E)|^2 + x^{0.5}\Psi_2^*(E)\Psi_1(E) - x^{0.5}\Psi_1^*(E)\Psi_2(E) + |\Psi_2(E)|^2 \big) +$$
$$+ \tilde{C}_1^*\tilde{C}_2 e^{-2iEt}\big( -\Psi_1^*(E)\Psi_1(-E) + x^{0.5}\Psi_2^*(E)\Psi_1(-E) - x^{0.5}\Psi_1^*(E)\Psi_2(-E) + \Psi_2^*(E)\Psi_2(-E) \big) +$$
$$\tilde{C}_2^*\tilde{C}_1 e^{2iEt}\big( -\Psi_1^*(-E)\Psi_1(E) + x^{0.5}\Psi_2^*(-E)\Psi_1(E) - x^{0.5}\Psi_1^*(-E)\Psi_2(E) + \Psi_2^*(-E)\Psi_2(E) \big) +$$
$$|\tilde{C}_2|^2 \big( -|\Psi_1(-E)|^2 + x^{0.5}\Psi_2^*(-E)\Psi_1(-E) - x^{0.5}\Psi_1^*(-E)\Psi_2(-E) + |\Psi_2(-E)|^2 \big) \Big)$$

$$\tag{12}$$

## 3. Main results and discussion

The dependence of the current on time according to (12) is presented in Figure 1.

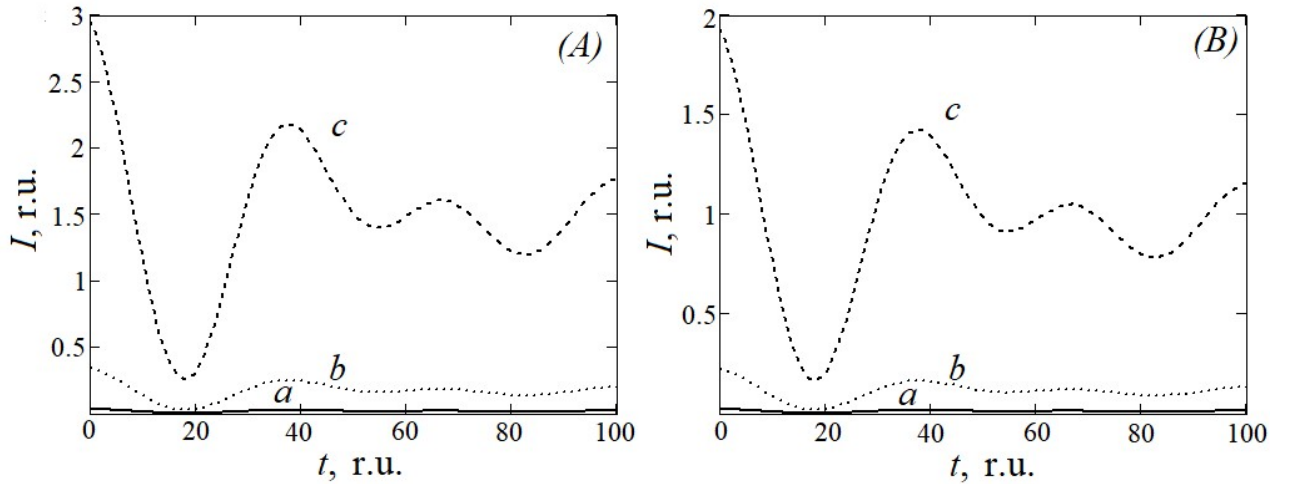

Fig.1. Dependence of current on time: (A) $d=4$, (B) $d=5$ ($l=1$): a) $r=2.1$; b) $r=2.2$; c) $r=2.3$. All values in the relatively units.

According to the Fig. 1 the magnitude of current oscillations during the ZB effect increases with distance from the event horizon. This can be attributed to the fact that the metric tends to be "flat" when away from the event horizon and the effects associated with curvature, which contribute to the interference of antiparticle-particles (one of the vivid explanations of the ZB effect), decrease. In fact, the dependence of the current on $r$ is more complex, and is shown in Fig. 2 for the selected point in time.

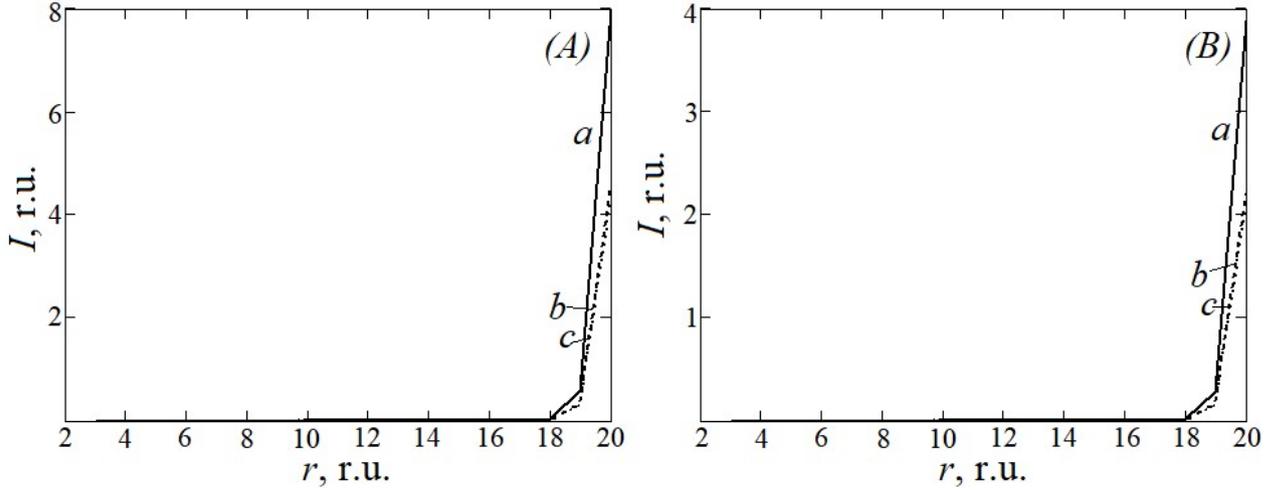

Fig.2. Dependence of the current on r: (A) $d=4$, (B) $d=5$, ($l=1$): a) $t=0$; b) $t=50$; c) $t=100$. All values in the relatively units.

As we observed earlier, the current increases the further we are from the event horizon. Note also that, all other parameters being equal, the value of $l$, according to (1), determines the position of the black hole's horizon. The corresponding dependences on the parameter $l$ are shown in Fig. 3

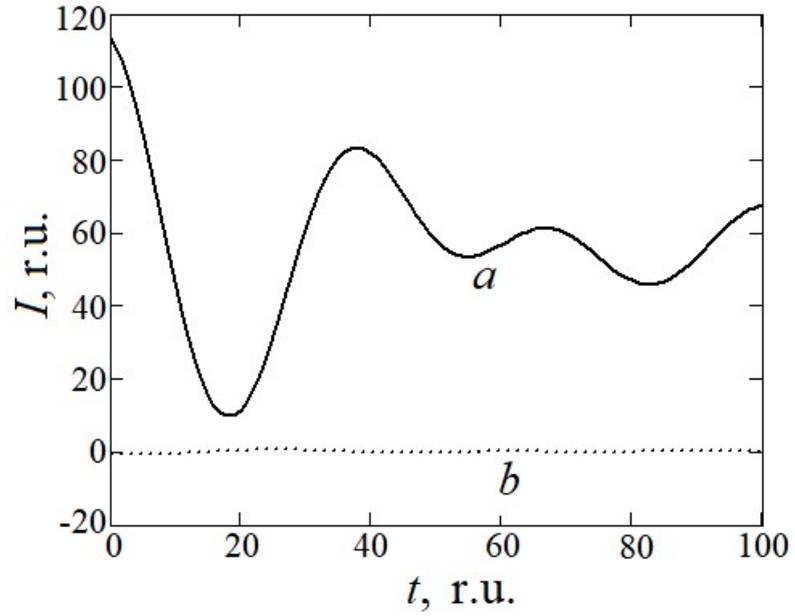

Fig.3. Dependence of current on time ($d=5$, $r=2.5$): a) $l=1$; b) $l=2$. All values in the relatively units.

It is experimentally possible to detect the ZB effect on radiation (13) from a black hole, whose dependence on time is shown in Figure 4:

$$Int = \frac{2q^2}{3c^3}\ddot{x}^2 \qquad (13)$$

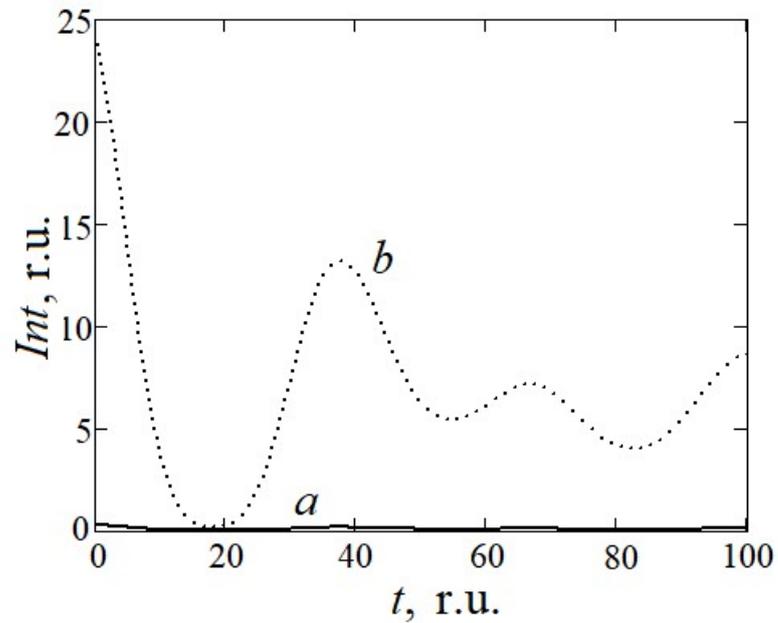

Fig.4. Dependence of radiation intensity on time ($d=5$, $l=1$): a) $r=2.1$; b) $r=2.2$. All values in the relatively units.

## 4. Discussion

Based on the calculations performed above, it can be concluded that the ZB effect for Chern-Simons 4 and 5 dimensional holes has the same regularities qualitatively. It differs only in the magnitude of the effect for different dimensions. We associate this with the appearance of the metric (1), where the division into radial and "angular" parts is made. Our research mainly focused on the dependence of the current on the radial coordinate, which is determined by the function $f(r)$, for which the dimension of space is a parameter. In accordance with the form of the function $f(r)$, the magnitude of the ZB effect decreases with increasing space dimension. We also note that the ZB effect substantially depends on the radial component, with an increase in which it increases too. In other words, as you move away from the event horizon, the current ZB increases. All this facts may be useful in the further study of quantum effects in the space of higher dimension.